\documentclass[12pt,preprint]{aastex}
\shorttitle{Fireball Heated by Neutrinos}
\shortauthors{Asano and Iwamoto}
\begin{document}
\title{Fireball Heated by Neutrinos}
\author{Katsuaki Asano and Shizuo Iwamoto}
\affil{Department of Earth and Space Science,
Osaka University, Toyonaka 560-0043, Japan}

\email{asano@vega.ess.sci.osaka-u.ac.jp, 
       iwamoto@vega.ess.sci.osaka-u.ac.jp}

\begin{abstract}
The fireball, the promising model of the gamma-ray burst (GRB),
is an opaque radiation plasma, whose energy is significantly
greater than its rest mass.
We numerically simulate the evolution of the fireball
heated by the neutrino-antineutrino annihilation process
for the spherically symmetric case.
We also derive analytical energy and momentum deposition rates
via neutrino scattering with thermalized electron-positron pairs
in the fireball.
In our simulation the matter is provided around the neutrinosphere
before neutrinos start to be emitted, and
the energy is injected
during a finite period of time $t_{\rm dur}$.
In the acceleration regime the matter shell is pushed from behind
by radiation pressure.
The Lorentz factor of the shell reaches the maximum value $\eta$ at
$r \simeq \eta^2 c t_{\rm dur}$.
After the fireball enters the coasting regime,
the velocity distribution in the shell becomes very flat.
The shell expansion rate $d W/dr$ can be much smaller than $\eta^{-2}$.
The runaway of temperature of the fireball
due to neutrino scattering with electron-positron pairs
does not occur in most cases.
The energy deposition due to scattering is not so significant.
\end{abstract}


\section{INTRODUCTION}

\indent

The relativistic fireball \citep{she90,ree92,mes93,sar95,sar96}
is  one of the most promising models of gamma-ray bursts (GRBs).
The sudden release of a large quantity of gamma-ray photons
into a compact region produces a fireball
that is an opaque radiation plasma whose energy is significantly
greater than its rest mass.
\citet{pir93} and \citet{mesl93} investigated the hydrodynamics of fireballs
and established basic behaviour in their evolution.
Most of the matter and energy is concentrated within a narrow shell.
In the radiation-dominated phase,
the rest mass and radiation energy density in the fireball behave as 
$\rho \propto r^{-3}$ and $e \propto r^{-4}$
($r$ is the location of the shell), respectively.
The bulk Lorentz factor $\gamma$ increases in proportion to $r$.
After the fireball enters the matter-dominated regime, 
as long as the shell is optically thick 
to scattering, the densities behave as $\rho \propto r^{-2}$ and $e \propto r^{-8/3}$,
and the shell eventually coasts with $\gamma =\eta \equiv E/Mc^2$,
where $M$ and $E$ are the total baryon rest mass and the total energy
of the fireball, respectively.
A simple analytical estimate and hydrodynamical simulations
indicate that the shell width expands as $W \simeq r/\eta^2$ after the
comoving energy density drops below the baryon rest mass density,
while the shell width remains constant during the radiation-dominated phase.
The actual behaviour of the fireball may be more complex depending
on the energy injection process of the central engine.
A detailed study of the fireball in various situations
will be important to determine the character of the central engine
in the future.

The rapid temporal variabilities observed require that the GRB
itself must arise from internal shocks within the flow.
The shortest time scale is about 1 msec \citep{bha92}.
The internal shock is generated  by the collision of two
shells with different speeds arising from inhomogeneity of the fireball.
In this paper we assume that
the inhomogeneity in the fireball can be represented by
multiple fireballs arising intermittently.

Notwithstanding the very high energy phenomenon ($\simeq 10^{52}$-$10^{54}$ ergs
for spherical symmetry), the baryon
density in the fireball must be extremely small.
In order to reproduce GRBs the final Lorentz factors
of shells should be $100$-$1000$.
This is the famous baryon contamination problem and still remains unsolved.
The central engine of GRBs is still beyond deep mist.
The source of GRBs may be one super massive
(failed) supernovae  \citep{woo93,pac98}
or may be a merger of two neutron stars or of a neutron star and a black hole
\citep{eic89,nar92,mes92,kat97,ruf98,ruf99}.

In these systems the hot core or accretion disk can emit neutrinos and antineutrinos.
The neutrino-antineutrino annihilation
into electrons and positrons
(hereafter neutrino pair-annihilation)
is a possible and important candidate to explain the energy source of GRBs
\citep{pac90,mes92b,jan96,ruf97,ruf98,ruf99,sal01}.
The central engine releases energy in a finite period of time at least
longer than $L/c$ where $L$ is the scale of the central engine.
Motivated by the delayed explosion of Type II supernovae,
the energy deposition rate due
to neutrino pair-annihilation above the neutrinosphere
has been calculated \citep{goo87,coo87,ber87}.
The energy deposition rate is proportional to $r^{-8}$ for large $r$,
and almost all deposition occurs near the neutrinosphere.
In hydrodynamic simulations of neutron star mergers,
\citet{ruf99} showed that neutrino pair-annihilation deposits 
energy in the vicinity of the torus at a rate of $(3$-$5)\times 10^{50}$ ergs ${\rm s^{-1}}$.
In the failed supernovae model \citet{mac99} predict that
the total energy deposited along the rotational axes of Kerr black holes is
$(1$-$14)\times 10^{51}$ ergs.

Most of the neutrinos' energy ($\gtrsim 99$ \%)
will escape to infinity without annihilation.
Since GRBs are very high energy phenomena,
it is favorable that the energy of neutrinos
is transformed into the fireball much more efficiently.
\citet{asa00,asa01} showed that the gravitational effect
does not change the energy deposition rate dramatically.
There is also another possible process to transfer the neutrinos' energy
to the fireball.
In the fireball, high-energy photons produce a large number of
electron-positron pairs at the initial stage.
The large optical depth due to electron-positron pairs
leads to thermal equilibrium \citep{she90}.
The neutrino scattering with the thermalized electron-positron pairs
(hereafter neutrino scattering)
can inject energy into the fireball.
This possibility was already pointed out by \citet{woo93}.

In this paper we discuss the evolution of the fireball
heated by neutrinos and the formation of shells using a simple spherical model.
We do not deal with a specific model of the central engine.
We discuss only the general evolution of the fireball with energy injection
in a finite period of time and the effect of neutrino scattering.
While baryonic matter has been considered to be injected together
with radiation energy in general,
we assume here that the matter is provided around the central engine
before neutrinos start to be emitted.
In \S 2 we review the formulae for the energy and momentum
deposition rates via neutrino pair-annihilation
and derive the deposition rates via neutrino scattering.
In \S 3 we numerically simulate the fireball heated by neutrinos
to investigate the evolution of the fireball and the deposition rates
via neutrino scattering.
Finally, \S 4 is appropriated for the summary.

\section{ENERGY AND MOMENTUM DEPOSITION RATE}

\subsection{Pair Annihilation}

\indent

In this subsection we review the formulae for the energy and momentum
deposition rates due to neutrino pair-annihilation.
For simplicity, we assume that neutrinos are emitted isotropically
from the neutrinosphere of radius $R_{\nu}$.
The neutrino number density in the phase space,
$f_\nu d^3 p_\nu =n_\nu(\varepsilon_\nu) \varepsilon_\nu^2 d \varepsilon_\nu d \Omega_\nu$,
is assumed to be conserved along the neutrinos' trajectory.
This is a good approximation as long as the energy efficiency of
neutrino pair-annihilation is very small.
Defining $X \equiv [1-( R_{\nu}/r )^2]^{1/2}$,
\citet{goo87} derived the energy deposition rate
due to neutrino pair-annihilation as
\begin{eqnarray}
\left. \frac{dE}{c dt dV} \right|_{\rm pair}&=&
\frac{K_{\rm p} G_{\rm F}^2}{12 \pi^2 c^2 R_{\nu}^4}
\dot{N_\nu} \dot{N_{\bar{\nu}}} (\langle \varepsilon_\nu \rangle
\langle \varepsilon_{\bar{\nu}}^2 \rangle
+\langle \varepsilon_\nu^2 \rangle \langle \varepsilon_{\bar{\nu}} \rangle)
F_E(r),
\label{epair}
\end{eqnarray}
where
\begin{eqnarray}
F_E(r)
=(1-X)^4 (X^2+4X+5),
\label{pair}
\end{eqnarray}
and the Fermi constant $G_{\rm F}^2=5.29 \times 10^{-44} {\rm cm^2\, MeV^{-2}}$.
The total emmision rate of neutrinos, $\dot{N_\nu}$,
and the average energy of neutrinos, $\langle \varepsilon_\nu \rangle$, are written as
\begin{eqnarray}
\dot{N_\nu} \langle \varepsilon_\nu^k \rangle
=(4 \pi R_\nu^2) (\pi c) \int \varepsilon_\nu^{2+k} n_\nu(\varepsilon_\nu)
d \varepsilon_\nu.
\end{eqnarray}
If we simplify the energy distribution of neutrinos
by using the effective temperature, $T_{\rm eff}$,
we can express the total emmision rate as
\begin{eqnarray}
\dot{N_\nu} \langle \varepsilon_\nu \rangle
=\frac{7 \pi^6 R_\nu^2}{30 h^3 c^2} T_{\rm eff}^4, \qquad
\dot{N_\nu} \langle \varepsilon_\nu^2 \rangle
=\frac{90 \pi^2 \zeta(5) R_\nu^2}{h^3 c^2} T_{\rm eff}^5,
\end{eqnarray}
where $\zeta(n)$ is the Riemann zeta function.
The constant $K_{\rm p}$ is written by
the Weinberg angle $\sin^2{\theta_{\rm W}}=0.23$.
To be exact we should sum up contributions of all species of neutrino.
Each species may be emitted at different emission rates.
When $\nu_{\rm e}$ and $\bar\nu_{\rm e}$ are dominant components,
$K_{\rm p}=0.124$.
If $\dot{N_\nu} \langle \varepsilon_\nu \rangle$
and $\dot{N_\nu} \langle \varepsilon_\nu^2 \rangle$ of all the six species of neutrino
are the same, we can include the contributions of $\nu_\mu$ and $\nu_\tau$
by setting $K_{\rm p}=0.178$.
The fraction and mean energy of each species of neutrino
depend on the model of the central engine (see, e.g., \cite{ruf97}).
In most cases the effective $K_{\rm p}$ may be between $0.124$ and $0.178$
unless the emission rate of $\nu_\mu$ or $\nu_\tau$ is much larger
than the rate of $\nu_{\rm e}$.
For $r \gg R_{\nu}$, the energy deposition rate behaves as $\propto
T_{\rm eff}^9 r^{-8}$.
Most of the energy is deposited near to the surface of the neutrinosphere.
We also obtain the momentum deposition rate
in a similar way as
\begin{eqnarray}
\left. \frac{dP}{dt dV} \right|_{\rm pair}&=&
\frac{K_{\rm p} G_{\rm F}^2}{12 \pi^2 c^2 R_{\nu}^4}
\dot{N_\nu} \dot{N_{\bar{\nu}}} (\langle \varepsilon_\nu \rangle
\langle \varepsilon_{\bar{\nu}}^2 \rangle
+\langle \varepsilon_\nu^2 \rangle \langle \varepsilon_{\bar{\nu}} \rangle)
F_P(r),
\end{eqnarray}
where
\begin{eqnarray}
F_P(r)
=(1-X)^4 (1+X) (3 X^2+9 X+8)/4.
\end{eqnarray}

\subsection{Neutrino Scattering}

In the fireball
high-energy photons produce a large number of
electron-positron pairs.
The large optical depth due to electron-positron pairs
leads to thermal equilibrium \citep{she90}.
Neutrino scattering with thermalized electron-positron pairs
can transfer the neutrinos' energy to the fireball.
We derive the energy and momentum deposition rates
through neutrino scattering in this subsection.
The electron-positron pairs are
assumed to be relativistic $\varepsilon_{\rm e} \gg m_{\rm e} c^2$.
For the reaction $\nu_{\rm e}+{\rm e}^{-} \to \nu_{\rm e}+{\rm e}^{-}$,
the differential cross-section is given by
\begin{eqnarray}
\frac{d \sigma}{dy}
=\frac{G_{\rm F}^2 (p_{i,{\nu}} \cdot p_{i,{\rm e}})}{2 \pi}
\left[ (2 \sin^2{\theta_{\rm W}}+1)^2+4\sin^4{\theta_{\rm W}}
(1-y)^2 \right],
\end{eqnarray}
where $y \equiv p_{i,{\rm e}} \cdot (p_{i,{\nu}} - p_{f,{\rm e}})
/(p_{i,{\nu}} \cdot p_{i,{\rm e}})$ is Lorentz invariant,
and $(p_{i,{\nu}} \cdot p_{i,{\rm e}})$ expresses the inner product of
the 4-momenta of a neutrino and an electron.
The subscripts ``$i$'' and ``$f$'' denote the initial and final states, respectively.
We can write the cross-sections for the other five species of neutrino
in a similar way (see, e.g., \citet{kim93}).

In the comoving frame of the relativistic fireball,
the energy of the neutrinos redshifts as $\varepsilon_\nu'=
\varepsilon_\nu/\gamma(1+\beta \cos{\theta'_\nu})$,
where $\gamma=1/(1-\beta^2)^{1/2}$ is the Lorentz factor of the fireball
and $\theta'_\nu$ is the incident angle of neutrinos to the $r-$axis
(see Fig. 1).
Hereafter values with a prime are measured in the comoving frame.
We have two scattering angles for a given value of $y$.
The energy a neutrino loses per one scattering
depends on the final scattering angle.
The average energy an electron/positron obtains for a given value of $y$
becomes
$\overline{\Delta \varepsilon'}=y (\varepsilon'_\nu-\varepsilon'_{\rm e})$.
The average momentum component an electron/positron obtains,
which is parallel to the neutrino's momentum, becomes
$c \overline{\Delta p'_\parallel}
=y (\varepsilon'_\nu-\varepsilon'_{\rm e} \cos{\hat{\theta'}})$,
where $\hat{\theta'}$ is the incidence angle between the two particles.
Here we assume that the random motion of electron-positron pairs is isotropic
in the fireball frame.
When a neutrino interacts with electrons distributed isotropically,
the mean vertical momentum the neutrino loses per one scattering is zero.
As we have assumed in \S 2.1, if $n_\nu(\varepsilon_\nu)$
of all the six species of neutrino
are the same, the energy a scattered electron (positron) gains
and the total cross section are averaged as
\begin{eqnarray}
\sum_s n_{\nu s}(\varepsilon'_\nu) \overline{\sigma_s \cdot \Delta \varepsilon'}
&=&n_\nu(\varepsilon'_\nu) 
\int_0^1 \sum_s \frac{d \sigma_s}{dy} y (\varepsilon'_\nu-\varepsilon'_{\rm e}) dy \\
&=& \frac{7}{12} n_\nu(\varepsilon'_\nu)
K_{\rm s} G_{\rm F}^2 (p'_{i,{\nu}} \cdot p'_{i,{\rm e}})
(\varepsilon'_\nu-\varepsilon'_{\rm e}),
\end{eqnarray}
where the summation is over all species of neutrino, and
$K_{\rm s} \equiv (24 \sin^4{\theta_{\rm W}}-4 \sin^2{\theta_{\rm W}}+3)/2 \pi
=0.533$.
When $\nu_{\rm e}$ and $\bar\nu_{\rm e}$ are dominant components,
$K_{\rm s}$ should be altered as
$(16 \sin^4{\theta_{\rm W}}+8 \sin^2{\theta_{\rm W}}+2)/4 \pi=0.373$.
To be more precise we have to estimate the deposition rate corresponding to
the variation of the distributions of the six species of neutrino.
We include the variation by adjusting $K_{\rm s}$ effectively
as we have done in \S 2.1.

The numbers of neutrinos and antineutrinos are assumed to be the same.
Then the reaction rates of electrons and positrons are the same because
of the symmetric property.
Since the energy distribution of the positrons is the same as
the distribution of the electrons in thermal equilibrium,
we include the contribution of positrons by doubling
the energy distribution of electrons $n_{\rm e}(\varepsilon'_{\rm e})$.
The energy deposition rate in the comoving frame is written as
\begin{eqnarray}
\left. \frac{dE}{c dt dV} \right|_{\rm scat}'&=&
\int  \int \int \int d \varepsilon'_\nu d \varepsilon'_{\rm e}
d \Omega'_\nu d \Omega'_{\rm e} \sum_s \varepsilon'^2_{\nu}
\varepsilon'^2_{\rm e} n_{\nu s}(\varepsilon'_\nu) n_{\rm e}(\varepsilon'_{\rm e})
\overline{\sigma_s \cdot \Delta \varepsilon'}
c (1-\cos{\hat{\theta'}}) \\
&=&
\frac{14 c K_{\rm s} G_{\rm F}^2 \pi^2}{3}
\int \int d \varepsilon'_\nu d \varepsilon'_{\rm e}
n_\nu(\varepsilon'_\nu) n_{\rm e}(\varepsilon'_{\rm e})
\varepsilon'^3_{\nu} \varepsilon'^3_{\rm e}
(\varepsilon'_\nu-\varepsilon'_{\rm e}) \nonumber \\
&\times& \int^{\theta'_{\rm max}}_0 d \theta'_\nu \sin{\theta'_\nu}
\int_0^\pi d \hat{\theta'} \sin{\hat{\theta'}} (1-\cos{\hat{\theta'}})^2,
\end{eqnarray}
where $\cos{\theta'_{\rm max}} \equiv (X-\beta)/(1-\beta X)$, including
the redshift effect.
Assuming the electron-positron pairs are in thermal equilibrium and
their temperature $T_{\rm e} \gg m_{\rm e} c^2$, we get
\begin{eqnarray}
\left. \frac{dE}{c dt dV}' \right|_{\rm scat}= f(T_{\rm e}(r)) G'_E(T_{\rm e},\beta;r),
\end{eqnarray}
where
\begin{eqnarray}
f(T_{\rm e}(r))
&=&\frac{7}{R_{\nu}^2 c} \frac{K_{\rm s} G_{\rm F}^2}{(h c)^3}
T_{\rm e}^4(r) \dot{N_\nu},
\end{eqnarray}
and
\begin{eqnarray}
G'_E(T_{\rm e},\beta;r)
&=&\gamma^2 (1-X) \left\{ \gamma \frac{7 \pi^4}{135} \langle \epsilon_\nu^2
\rangle
\left[1-\frac{3}{2} \beta (1+X)+ \nonumber \right. \right. \\
&& \left. \beta^2
(1+X+X^2) -\frac{\beta^3}{4} (1+X) (1+X^2) \right] \nonumber \\
&& \left. -20 \zeta(5) \langle \epsilon_\nu \rangle
T_{\rm e}(r) \left[ 1-\beta (1+X)+\frac{\beta^2}{3} (1+X+X^2) \right] \right\}.
\end{eqnarray}
The temperature $T_{\rm e}(r)$ is measured in the comoving frame.
The momentum deposition rate can be obtained in a similar fashion,
using the parallel component of momentum to the $r$-axis,
$\cos{\theta'_\nu} \overline{\Delta p'_\parallel}$.
Carrying out the Lorentz transformation, we obtain
the energy deposition rate in the observer frame as
$dE/c dt dV|_{\rm scat}=f(T_{\rm e}(r)) G_E(T_{\rm e}, \beta;r)$,
where
\begin{eqnarray}
G_E(T_{\rm e},\beta;r)
=\gamma \left\{ \gamma \frac{7 \pi^4}{135} \langle \epsilon_\nu^2
\rangle
\left[1-\beta (1+X)+\frac{\beta^2}{3}
(1+X+X^2) \right] \nonumber \right. \\
-20 \zeta(5) \langle \epsilon_\nu \rangle
T_{\rm e}(r) \left[ \frac{1}{2} (3 \gamma^2-1)+\frac{\beta}{4}
(1-6 \gamma^2) (1+X) \right. \nonumber \\
\left. \left.+\frac{1}{2} (\gamma^2-1) (1+X+X^2) \right] \right\}
(1-X).
\end{eqnarray}
In the case $T_{\rm e}(r) \lesssim m_{\rm e} c^2$ these formulae overestimate the rates.
These equations give the upper limit of the energy and momentum deposition rates.
For $\beta \simeq 0$ and $r \gg R_{\nu}$, introducing the effective temperature of
the neutrinos,
the energy deposition rate behaves as $\propto (T_{\rm eff}^5 T^4_{\rm e}(r)
-T^5_{\rm e}(r) T_{\rm eff}^4) r^{-2}$.
The deposition rate depends on the temperature of the fireball.
When $r=R_\nu$ and $\beta=0$,
the constant factor, except for the temperature dependence,
in the deposition rate via neutrino scattering
is larger than the factor via pair-annihilation by a factor of $\sim 5$.
If $T_{\rm e} \simeq T_{\rm eff}$,
the energy injection rate for neutrino scattering
becomes comparable to the rate for neutrino pair-annihilation.
However, if $T_{\rm e} \geq T_{\rm eff}$ the scattering effect
cools down the fireball inversely.
The deposition rate via pair-annihilation behaves as $\propto
r^{-8}$.
If $T^4_{\rm e}(r)$ has a flatter distribution than $r^{-6}$,
much more energy could be injected by scattering than
energy injected by annihilation at a great distance.

The momentum deposition rate in the observer frame
is obtained
by $dP/dt dV|_{\rm scat}=f(T_{\rm e}(r)) G_P(T_{\rm e},\beta;r)$, where
\begin{eqnarray}
G_P(T_{\rm e},\beta;r)
=\gamma \left\{ \gamma \frac{7 \pi^4}{135} \langle \epsilon_\nu^2
\rangle \left[\frac{1}{2} (1+X)
-\frac{2}{3} \beta (1+X+X^2)+\frac{\beta^2}{4} (1+X) (1+X^2) \right] \nonumber \right. \\
\left. +10 \zeta(5) \langle \epsilon_\nu \rangle
T_{\rm e}(r) \left[ \left(3 \gamma^2-\frac{5}{2} \right) (1+X)
-3 \beta \gamma^2-\frac{\beta \gamma^2}{3} (1+2 \beta^2) (1+X+X^2) \right] \right\}
(1-X).
\end{eqnarray}
For $\beta \simeq 0$, neutrinos accelerate the fireball
regardless of $T_{\rm e}$.
Neutrino scattering can decelerate the flow for large $\gamma$
like Compton dragging.

There is a possibility that the temperature of the fireball could run away
via neutrino scattering \citep{woo93}
and reach $T_{\rm e} \simeq T_{\rm eff}$.
In addition the total energy could be enhanced by the energy deposition
at large $r$ if $T_{\rm e}(r)$ distributes flatter than $r^{-6}$.
However, it seems difficult to lead $T_{\rm e}$ to the runaway,
because the dependence $\sim T^4_{\rm e}$
makes the deposition rate due to neutrino scattering much weaker
than the rate due to neutrino pair-annihilation giving a slightly smaller $T_{\rm e}$
than $T_{\rm eff}$.
To reach higher temperatures the expansion speed of the plasma
should be slow enough during the energy injection such that
the heating effect via neutrinos is dominant rather than adiabatic cooling.
In most cases we need to simulate the evolution of the fireball numerically
in order to estimate the conclusive energy deposition via neutrinos.

\section{NUMERICAL CALCULATION}

In this section we numerically simulate the fireball
heated by neutrinos.
Neutrinos are assumed to be emitted isotropically from the neutrinosphere
during a finite period of time.
As \citet{pir93} assumed in their calculation,
we assume the fireball is optically thick.
The radiation and matter at each radius behave like a single fluid
moving with the same velocity.
Since the radiation pressure dominates,
the pressure $p$ and the energy density $e$ of radiation and relativistic
electron-positron pairs are related by $p=e/3$.
The relativistic conservation equations of baryon number, energy
and momentum are written by
\begin{eqnarray}
\frac{1}{c} \frac{\partial}{\partial t} \left( \rho \gamma \right)
=-\frac{1}{r^2} \frac{\partial}{\partial r} \left( r^2 \rho u
\right),
\label{number}
\end{eqnarray}
\begin{eqnarray}
\frac{1}{c} \frac{\partial}{\partial t} \left(
\frac{4}{3} e \gamma^2-\frac{e}{3}+\rho \gamma^2 \right)
=-\frac{1}{r^2} \frac{\partial}{\partial r} \left[ r^2 \gamma u
\left(\frac{4}{3} e+\rho \right) \right] +\frac{d E}{c dt dV},
\label{energy}
\end{eqnarray}
\begin{eqnarray}
\frac{1}{c} \frac{\partial}{\partial t} \left[
u \gamma \left( \frac{4}{3} e+\rho \right) \right]
=-\frac{1}{r^2} \frac{\partial}{\partial r} \left[ r^2 u^2
\left(\frac{4}{3} e+\rho \right) \right]-
\frac{1}{3} \frac{\partial e}{\partial r} +\frac{d P}{dt dV},
\label{momentum}
\end{eqnarray}
where $u \equiv \gamma \beta$.
The rest mass energy density of baryons $\rho$ and the non-baryonic energy
density $e$ are measured in the comoving frame.

We use the relativistic ``Roe-solver'' method
in the computation of one-dimensional relativistic flow.
The {\it Riemann problem} is approximated numerically
by means of a specially chosen linearized form,
requiring less computational effort.
This Eulerian code was developed for the relativistic case
by \citet{mel91}.
However, our computation is not so complicated as the computation of Mellema et al.
(1991), because we assume spherical symmetry and neglect the gravitational effect.
Our code is relatively simple and has passed several standard tests.
In addition we have checked our numerical code
by reproducing the results of the simulation in \citet{pir93}.
We adopt the spatial and time resolution as $\Delta r=R_\nu/3000$
and $c \Delta t=\Delta r/10$, respectively.
In order to discuss the evolution of shell width, however,
we need to simulate with a very high spatial resolution.
The expansion/contraction rate of shell width may be limited
as $|d W/dr| \lesssim 1/\gamma^2$.
It will take very long time to follow the change of shell width accurately
for large $\gamma$ because of the high resolution.
In our simulation, judging from test calculations,
the expansion/contraction rate is not so reliable for $\gamma \gtrsim 10$.
However the rate of change is small enough for large $\gamma$, in our calculation,
such that we can follow the approximate behaviour of the fireball.

First we investigate the evolution of the fireball
in an ideal situation to produce a relativistic flow.
Cold baryonic matter is provided around the neutrinosphere
before the simulation starts.
Then we study the effect of neutrino scattering
on the energy deposition rate.

\subsection{Evolution of the Fireball}

In order to produce a relativistic flow,
we assume an ideal situation.
The initial profile of the matter that we present is
compact as $\rho(r,t=0) \propto (R_0^8+r^8)^{-1}$ for $r \geq R_{\nu}$.
The matter is initially cold and static as $e(r,0)=0$ and $\gamma(r,0)=1$.
In this simulation the energy is injected at the constant rate, $\dot{E}$,
during $0 \leq t \leq (r-R_{\nu})/c+t_{\rm dur}$,
while in previous studies they
had provided radiation energy before the simulations started
\citep{pir93,mesl93}.
Spatial distribution of $\dot{E}$ is given by equation (\ref{pair}).
We neglect the effect of neutrino scattering in this subsection.
We set the initial core radius of the matter as $R_0=c t_{\rm dur}$ hereafter.
The rest mass energy is normalized as
$4 \pi \eta \int dr r^2 \rho =\dot{E} t_{\rm dur}$,
where $\eta$ is the ratio of the radiation energy to the rest mass energy.
The mean final Lorentz factor of the matter shell should be $\eta+1$.
Thus, the physical conditions in this simulation are
determined by the two parameters $\eta$
and $\tau_{\rm dur} \equiv c t_{\rm dur}/R_{\nu}$.

First of all we show results
for the case of $\eta=60$ and $\tau_{\rm dur} =3$ as an example.
If we introduce physical units as $R_\nu=10$ km and $T_{\rm eff}=10$ MeV,
the initial matter density at $r=R_\nu$ corresponds to
$2.4 \times 10^4$ g ${\rm cm^{-3}}$ in this case.
Figure 2-4 are the profiles of the fireball at different times.
At $\tau \equiv c t/R_{\nu}=3$ (Fig. 2),
namely, at the end of the energy injection,
the matter is swept out by the radiation pressure.
The information about the initial configuration of the matter is almost lost.
The approximation $\rho \ll e$, which has been assumed conventionally
in discussions about the initial evolution of the fireball,
already failed around the matter shell at this time.
The matter shell is compressed by the radiation pressure.
The FWHM of the matter shell in the observer frame
(the profile of $\rho \gamma$)
is about $0.1 R_{\nu}$ and remains nearly constant within the numerical error limit
hereafter.
The behavior of the fluid around the neutrinosphere agrees with
the stationary solution of pure radiation flow with energy injection.
We have confirmed that $e$ and $\gamma$ behind the matter shell in Figure 2
are very close to the numerical solutions
obtained by setting the left-hand side of equations (\ref{number})-(\ref{momentum})
to zero and $\rho=0$.
Figure 3 is the fireball at $\tau=6$.
The energy injection has already finished.
The pure radiation flow, which had been deposited later,
is chasing the matter shell.
Around the matter shell the forward and reverse shocks are formed.
The reverse shock is radiation dominant, while the forward one
is matter dominant.
In Figure 4 we display the fireball at $\tau=30$.
The Lorentz factor becomes a maximum behind the shell,
where the radiation is dominant.
The radiation flow pushes the matter from behind.
The matter shell itself accelerates more slowly than the prediction
by the simple analysis, $\gamma=(r/R_{\nu})$.
The behaviour of $u$ at the point where $\rho \gamma$ has the maximum value
is approximated as $u_{\rm M} \sim r^{0.5}$ at this moment.
The value $u_{\rm M}$ does not accelerate obeying a simple power law.
The power law index declines as $u_{\rm M} \propto r^{0.4}$,
$r^{0.3}$,... with increasing $r$.
We need a very long time to reach $u_{\rm M} = \eta$ in our computation.
The maximum values of $\rho$ and $e$ do not obey simple power-law behaviour.

In Figure 5 we plot $u_{\rm M}$ for different values of $\tau_{\rm dur}$ and $\eta=60$.
The acceleration time is roughly in proportion to
the duration time of the energy release.
When the shell reaches $r=\eta R_\nu$, $\gamma$ becomes
a few tens of a percent of $\eta$.
The maximum Lorentz factor in the whole system increases in proportion to $r$.
Extrapolating $u_{\rm M}$ from Figure 5, we guess
$u_{\rm M}$ may reach $\eta$ at $r \simeq \eta^2 \tau_{\rm dur} R_{\nu}$.
At the end of the energy injection the distance from $R_\nu$ to the head of the shell
is about $\tau_{\rm dur} R_\nu$.
Therefore, the energy injected at the end will catch up with the matter shell
of $\gamma \simeq \eta$ at $r \simeq \eta^2 \tau_{\rm dur} R_{\nu}$.
This is the reason $u_{\rm M}$ reaches $\eta$
at $r \simeq \eta^2 \tau_{\rm dur} R_{\nu}$.

To economize the calculation time we demonstrate the evolution of the fireball
for $\eta=10$ and $\tau_{\rm dur}=1$.
In Figure 6 we plot $u_{\rm M}$ and FWHM of the profile of $\rho \gamma$.
Since the initial matter distribution is narrow in comparison with the case
$\tau_{\rm dur} =3$, the radiation pressure expands the matter component initially.
Then the matter component is compressed.
Figure 7 shows that the rear part of the matter shell is faster than
the head part at $\tau=12$ that implies the shell is compressed.
The shell begins to expand again from $\tau \simeq 30$.
In this period the rear part of the matter shell
is slightly slower than the head part, as is shown in Figure 8.
The expansion rate is $d W/dr \simeq 10^{-3} < \eta^{-2}$,
which is close to the spatial resolution $\Delta r/R_\nu$ in our simulation.
This value may not be accurate because of
the size of the numerical error.
However, the difference of the Lorentz factor in the shell
is rather small as $\delta\gamma \ll \gamma$.
At $\tau=105$ the maximum and minimum $u$ within FWHM are
$10.28$ and $9.89$, respectively.
The maximum relative velocity within FWHM in the observer frame
is $3.8 \times 10^{-4} c$ that suggests $dW/dr \simeq 4 \times 10^{-4}$,
which is smaller than our error limit in the computation.
Thus, we can expect a smaller expansion rate than $\eta^{-2}$ by a factor of $10$ or $10^2$.
As we have assumed, $u_{\rm M}$ reaches $\eta$ at $r/R_\nu \simeq \tau \simeq \eta^2=100$.
We can expect that the behaviour of the fireball
for large $\eta$ would be similar to the above results.

We refer the simulation in \citet{pir93} to compare
with the case in which radiation energy has been given before the simulation starts.
In their simulation the initial profiles are
given as $e(r,0) \propto (R_0^8+r^8)^{-1}$ and $\rho(r,0)=e/\eta$.
According to Figure 2 in \citet{pir93}
the matter shell expands even in the acceleration regime.
At $\tau \simeq 100$ in our notation the shell width
grows to a few times the size of $R_0$.
The Lorentz factor in the matter shell increasingly distributes
with radius and
their difference in the shell is rather large as $\delta\gamma \simeq \gamma$.

The results of this subsection can be summarized as follows:
In the acceleration regime the shell is pushed from behind
by the radiation pressure.
The expansion of the shell is suppressed in this regime.
The acceleration declines as the shell evolves.
The Lorentz factor of the shell reaches $\eta$ at $r\simeq \eta^2 \tau_{\rm dur} R_{\nu}$.
After the fireball enters the coasting regime,
the shell begins to expand.
The velocity distribution in the shell is very flat.
We need a more accurate numerical simulation and longer calculation time
to estimate the expansion rate.
There is a possibility that the expansion rate is much less than $\eta^{-2}$.
The simulation in this subsection is merely one example of the evolution of the fireball
being affected by the energy deposition due to neutrinos.
However, when radiation fluid is deposited inside the initial matter distribution
during a finite period of time, the fireball may evolve in a similar manner
regardless of the details of the spatial distribution of $\dot{E}$ and initial matter.

\subsection{Effect of Neutrino Scattering}

In this subsection we simulate the fireball
including the effect of neutrino scattering.
Our purpose is to investigate the enhancement of the energy deposition
by the scattering effect.
The energy deposition by neutrinos including the scattering effect
is a nonlinear process.
The rate of neutrino scattering
depends on the temperature of the fireball.
As is shown in the former subsection,
the temperature around the neutrinosphere can be expressed by
the stationary solution of the pure radiation flow
as long as the initial matter density is sufficiently low.
It is easily found that the maximum temperature,
which is obtained from the stationary solution,
is much lower than the effective temperature of the neutrino.
In an ideal situation for producing the relativistic shell as in \S 3.1,
the effect of neutrino scattering is negligible.

However the environment is generally very dense around candidates of the central engine:
the central region of massive stars,
merging neutron stars, and so on.
In a very dense environment the matter cannot be swept out promptly.
The energy injected by neutrinos may stand near to the neutrinosphere,
and the fireball temperature becomes higher than the case
we considered in the former subsection.
We investigate the effect of neutrino scattering
for a dense environment.
In order to consider the most effective case on neutrino scattering,
we make the matter density constant as $10^9 {\rm g/cm^3}$
and set $R_\nu=10$ km and $T_{\rm eff}=15$ MeV.
Similarly, as in the former subsection, the energy deposition due to pair-annihilation
is given by equation (\ref{epair}) with $K_{\rm p}=0.178$.
We add the contribution of the scattering effect discussed in \S 2.2
to equations (\ref{number})-(\ref{momentum}) with $K_{\rm s}=0.533$.
The temperature of the fireball is obtained from
$T_{\rm e}^4=60 (\hbar c)^3 e/(11 \pi^2)$.
Since our purpose is to investigate the effect of scattering
between neutrinos and electron-positron pairs in the extreme environment,
we ignore the interaction between baryons and neutrinos.
As a matter of fact, for $T_{\rm eff}=15$ MeV,
the energy deposition rate due to the neutrino-baryon interaction
($\propto T_{\rm eff}^6 \rho$ for $\beta=0$)
becomes dominant rather than the rate due to pair-annihilation
at $r=R_\nu$, in a case in which the initial matter
density is larger than $7 \times10^9 {\rm g/cm^3}$.
As long as $\rho$ is constant, the total energy deposition via
the neutrino-baryon interaction is larger than the deposition via pair-annihilation.

In Figure 9, we display the fireball at $\tau=1$.
We have normalized the length scale and energy density
by 10 km and $(10^5$ g ${\rm cm^{-3}}) c^2$, respectively.
As we have assumed above, the expansion speed is slow
and the radiation energy stands inside the shell.
The distribution of $e$ inside the matter shell is nearly flat.
The typical temperature $T_{\rm e}$
inside the matter shell, where the radiation is dominant,
is about 5 MeV, smaller than the effective temperature of neutrinos.
At $r=R_\nu$ the scattering effect enhances the deposition rate by
only $5$ \%,
while the total enhancement is $20$ \% at $\tau=10$ (see Fig. 10).
This implies that
a considerable amount of energy is
deposited in a wide range of $r$ by scattering.

The temperature $T_{\rm e}(R_\nu)$ increases
after the start of the energy deposition (see Fig. 11).
However, the huge energy injection ($T_{\rm eff}=15$ MeV)
sweeps out the matter immediately.
Then adiabatic cooling
due to the expansion becomes stronger
than the neutrino heating.
The temperature $T_{\rm e}$ around the neutrinosphere
declines and reaches the value given by the stationary
solution.

If we enlarge the dimension of the region as $R_\nu=100$ km,
the energy deposition rate ($\propto R_\nu^3$)
and the energy efficiency ($\propto R_\nu$) due to pair-annihilation
increase.
The dynamical time scale becomes larger by a factor of 10,
while the energy injection rate per volume remains constant
for the given value of $r/R_\nu$.
Therefore, the injected energy per baryon during a dynamical time scale
increases, which leads to high temperatures.
In our case the fireball temperature $T_{\rm e}$ increases to $\sim 7$ MeV (Fig. 11).
The energy deposition is enhanced by 60\% because of scattering at $\tau=10$.
The qualitative behaviour in this case is almost the same as in the former case.

On the other hand, if the effective temperature  decreases to $T_{\rm eff}=5$ MeV,
the smaller energy injection cannot promptly sweep out the matter.
The adiabatic cooling effect is weaker
than the heating effect, initially.
The temperature $T_{\rm e}(R_\nu)$ increases with time,
but remains much smaller than $T_{\rm eff}$ (Fig. 11)
in comparison with the case $T_{\rm eff}=15$ MeV.
Then the increasing rate of $T_{\rm e}(R_\nu)$ is saturated.
To make matters worse,
the energy deposition due to the neutrino-baryon interaction
is practically dominant for lower $T_{\rm e}$.
Neutrino pair-annihilation and scattering play secondary roles.
Such a situation is not adequate for our purpose.
Reduction of the initial matter density leads to a prompt sweeping out of matter.
Thus, it is very difficult to make $T_{\rm e}$ large.

The temperature of the fireball does not run away in our cases
and the nonlinear effect is not so strong.
The energy deposition due to scattering is not so important
even under the extreme condition we adopted.

\section{SUMMARY}

We have simulated the fireball in which radiation energy
is deposited in a narrow region ($\dot{E} \sim r^{-8}$)
during a finite period of time.
The baryonic matter is initially prepared around the neutrinosphere.
In the acceleration regime the matter shell is pushed from behind
by the radiation pressure.
The Lorentz factor of the shell reaches $\eta$ at
$r \simeq \eta^2 \tau_{\rm dur} R_{\nu}$.
\citet{mes00} estimated that wind becomes optically thin
at $\sim 10^6 R_\nu$ with typical parameters.
If $\tau_{\rm dur} \gg 1$,
the shell can be optically thin before $\gamma$ reaches $\eta$
even if $\eta$ is smaller than the critical value $\eta_*$
(see \citet{mes00}).
The maximum values of $\rho$ and $e$ do not obey a simple power-law behavior.
After the fireball enters the coasting regime,
the shell may begin to expand.
The velocity distribution in the shell is rather flat
in comparison with the case
in which the radiation energy has been given before the simulation starts.
Although we need a more accurate numerical simulation
to estimate the expansion rate quantitatively,
$d W/dr$ may be smaller than $\eta^{-2}$.

The short time scale of the temporal profile of GRBs
requires that the shell width should remain thin at the collision radius.
If the shell width expands as $W=r/\eta^2$,
the width becomes $\sim 10^{10}$ cm for $\eta=100$
at the typical collision radius $\sim 10^{14}$ cm.
Then the observed emission time scale of one collision
becomes $\sim 1$ sec, which is longer than the observed timescale of the pulse rise.
In addition, if the shell expands, the magnetic field in the shell
may weaken.
It is very difficult to reproduce hard spectra of observed GRBs in this case
\citep{gue01}.
Therefore, it is favorable that the shell width is constant.
The actual shell expansion depends on the Lorentz factor distribution
across the shell, which is determined by the energy ejection process
around the central engine.
Although the actual activity of the central engine is more complex than our case,
our simulation shows a possibility that the expansion rate $dW/dr$ is smaller
than $\eta^{-2}$.
Our simulation is one simple example of the evolution of the fireball.
If baryonic matter is also injected with radiation energy,
the behaviour of the fireball may be close to the behaviour
in \citet{pir93}.
In our case baryonic matter should be injected
during the quiescent period of the central engine.
The matter could be injected from the side if the central engine has a jetlike structure.
But we do not know if this is realistic or not.
In the future, we will have to examine whether the shell width remains narrow
in some kind of model of the central engine.
Future observations may constrain shell's structure
and give a clue to determining the central engine.

As for the effect of neutrino scattering,
the temperature of the fireball does not run away in most cases
that we assume,
and the nonlinear effect is not so strong.
Although the scattering effect can enhance the energy deposition rate
by a few tens of a percent,
the effect on the deposition rate is not so important
even for the extreme conditions we assumed.
When other processes of energy injection,
like magnetic reconnection, etc., provide a large number of pairs,
the effect of neutrino scattering can be larger.
In any case a very dense environment is required for raising the
fireball temperature high enough in a wide range of $r$.
Under the dense environment the baryon contamination is crucial,
while the higher temperature increases the energy deposition due to the scattering effect.
Since the energy enhancement due to the scattering effect is not significant,
it is difficult to make $\gamma$ large enough
in the situation where the effect of neutrino scattering
becomes important.

\acknowledgments

We thank Dr. A. Iida for his useful advice concerning the hydrodynamical simulation.
The authors are supported by a Research Fellowship of the Japan Society for
the Promotion of Science.

\clearpage

\begin{figure}
\plotone{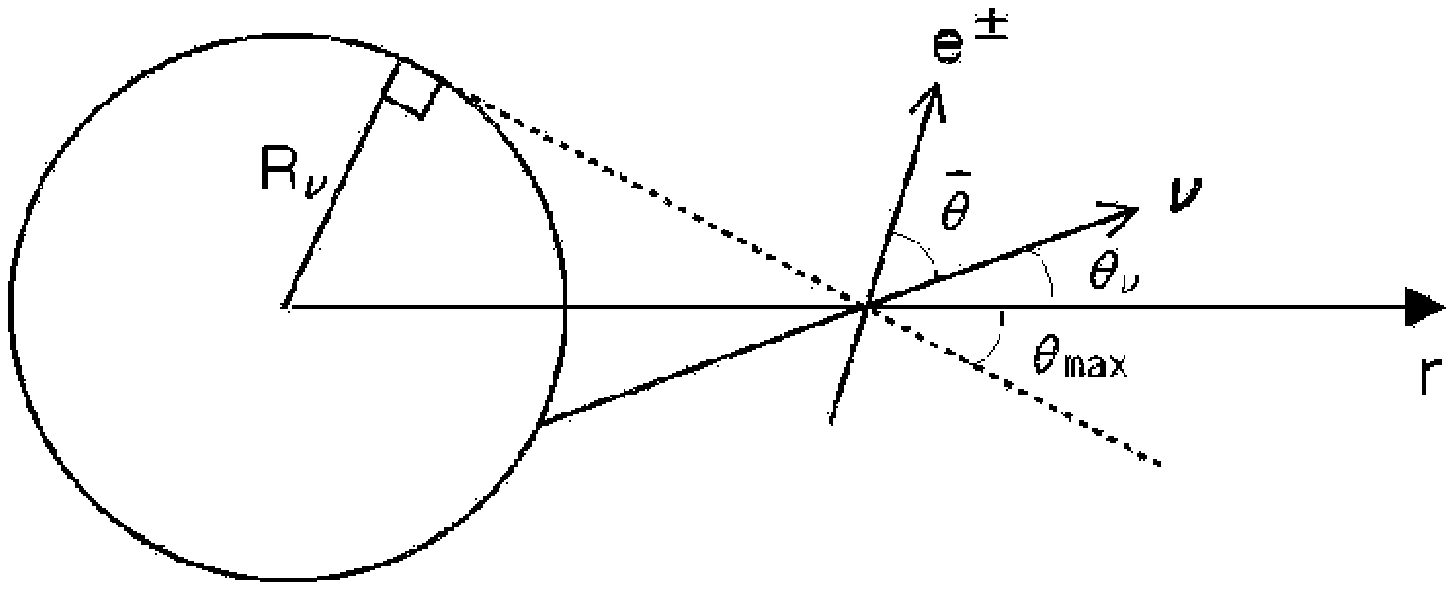}
\caption{Definitions of $\theta_{\rm max}$, $\theta_\nu$, and $\hat{\theta}$.}
\end{figure}

\clearpage

\begin{figure}
\plotone{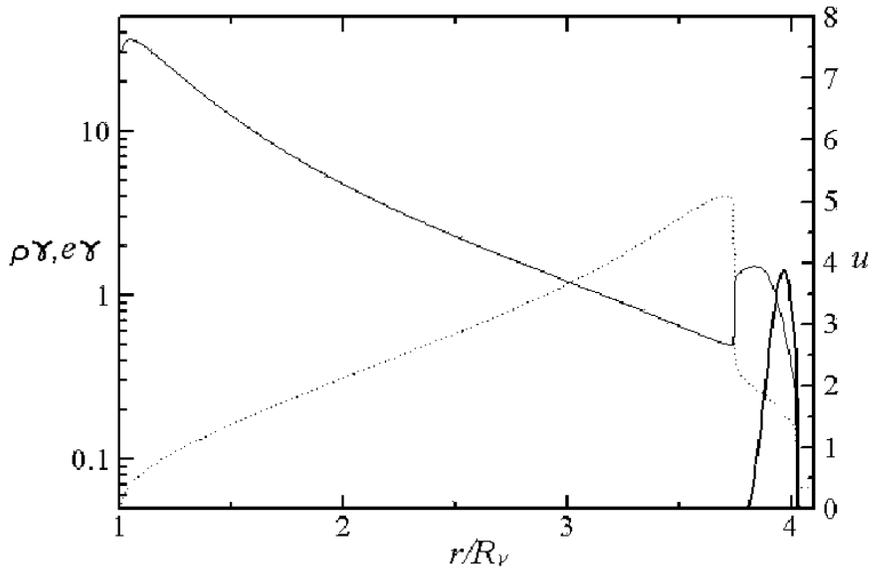}
\caption{Fireball at $\tau=3$ for $\eta=60$ and $\tau_{\rm dur}=3$.
The thick and thin solid lines are $\rho \gamma$ and $e \gamma$, respectively.
The dotted line is $u \equiv \gamma \beta$;
$\rho \gamma$ and $e \gamma$ are plotted in logarithmic scale.
Here we neglect the effect of neutrino scattering.}
\end{figure}

\clearpage

\begin{figure}
\plotone{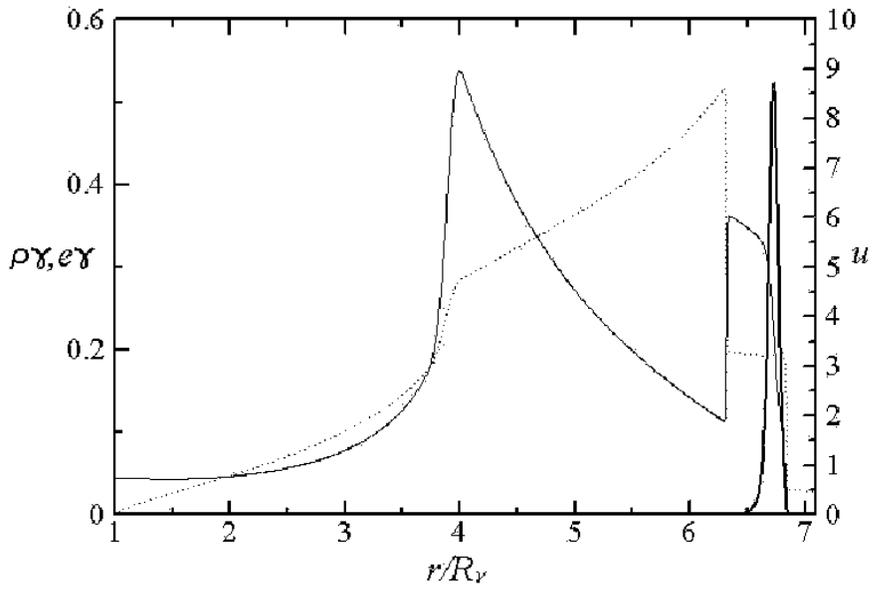}
\caption{Same as Fig.2, but for $\tau=6$;
$\rho \gamma$ and $e \gamma$ are plotted in linear scale}
\end{figure}

\clearpage

\begin{figure}
\plotone{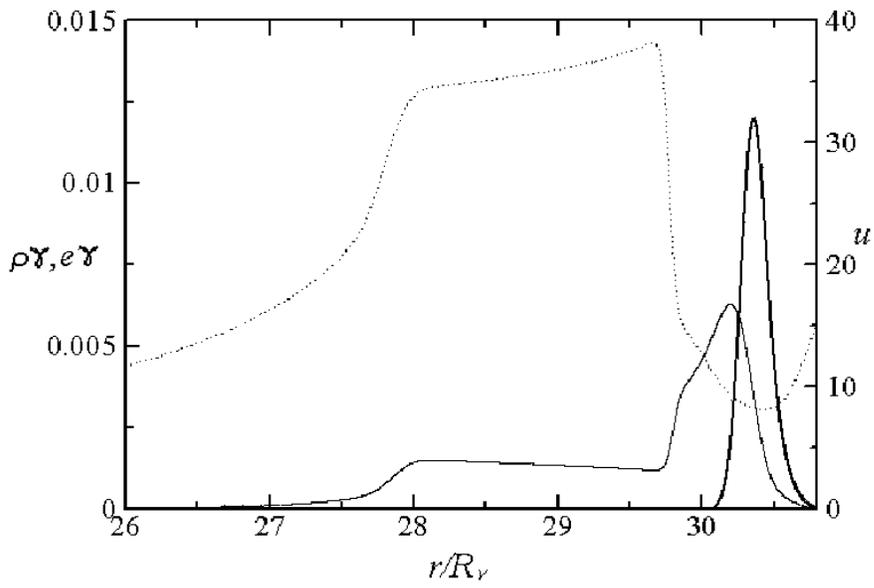}
\caption{Same as Fig.3, but $\tau=30$}
\end{figure}

\clearpage

\begin{figure}
\plotone{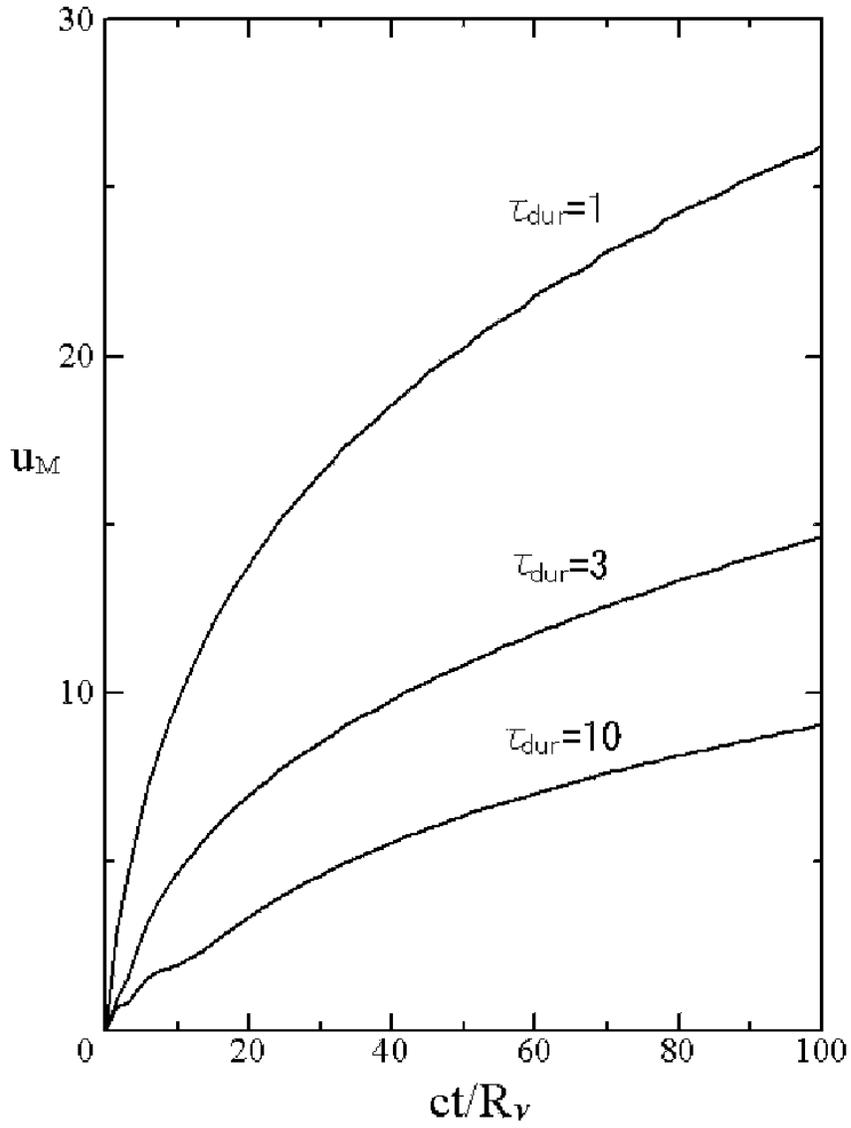}
\caption{Values of $u_{\rm M}$:
the value $u$ at the maximum point of $\rho \gamma$ for $\eta=60$}
\end{figure}

\clearpage

\begin{figure}
\plotone{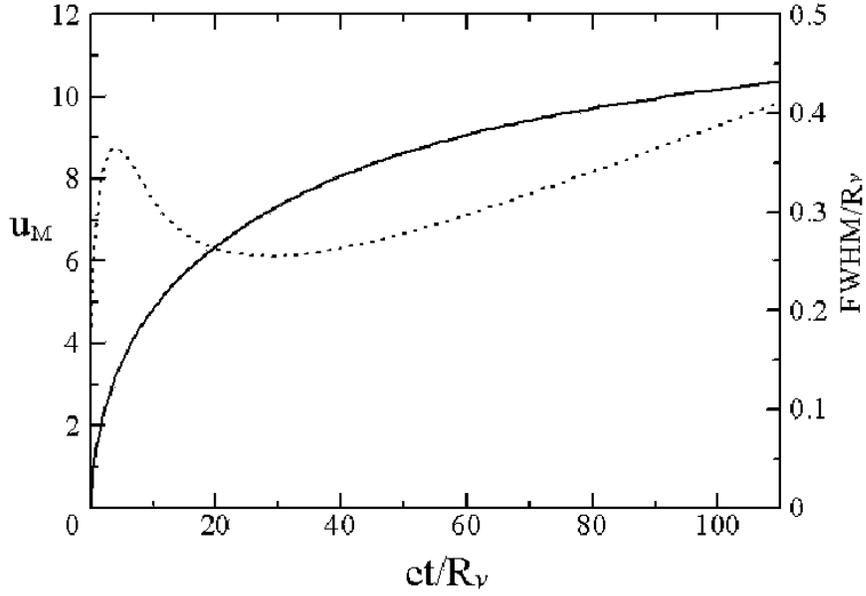}
\caption{Values of $u_{\rm M}$ (solid line) and
FWHM of the profile of $\rho \gamma$  (dotted line)
for $\eta=10$ and $\tau_{\rm dur}=1$}
\end{figure}

\clearpage

\begin{figure}
\plotone{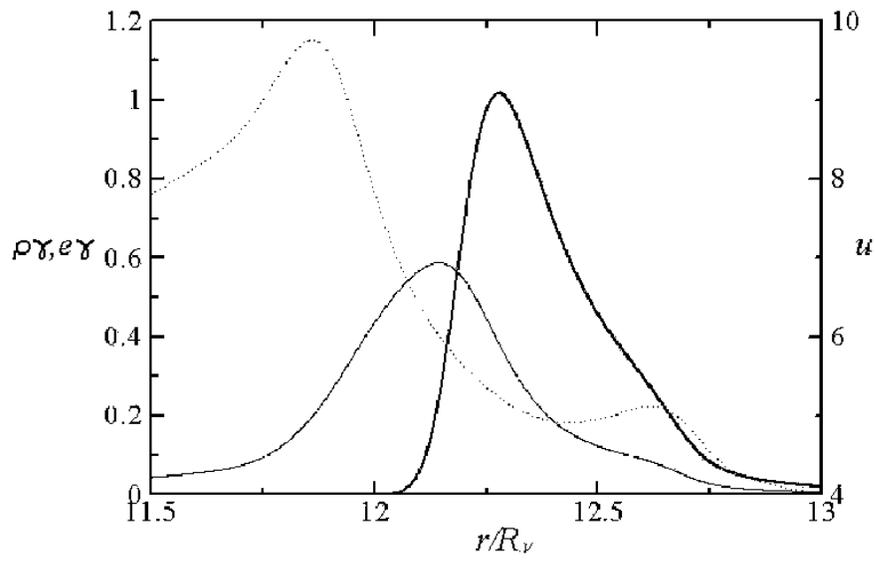}
\caption{The fireball at $\tau=12$ for $\eta=10$ and $\tau_{\rm dur}=1$}
\end{figure}

\clearpage

\begin{figure}
\plotone{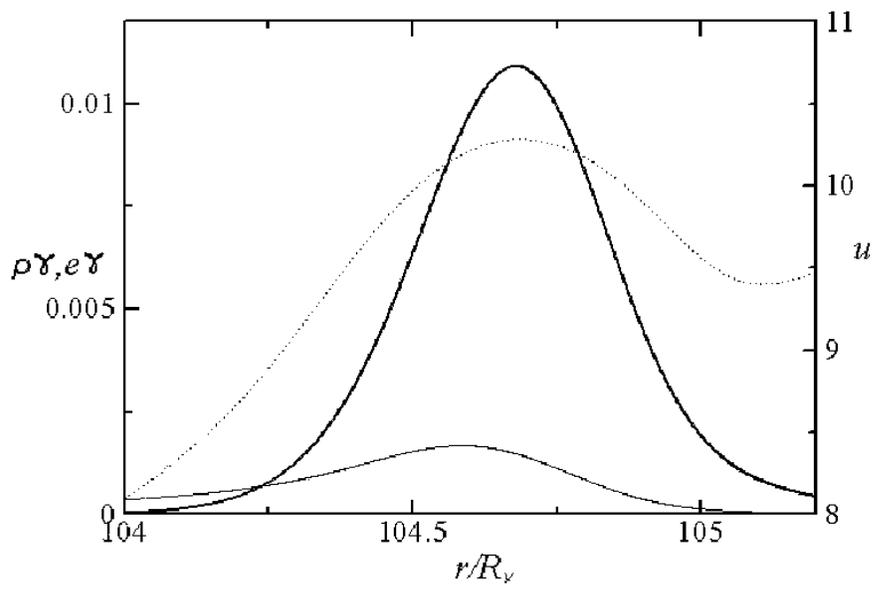}
\caption{Same as Fig.7, but $\tau=105$}
\end{figure}

\clearpage

\begin{figure}
\plotone{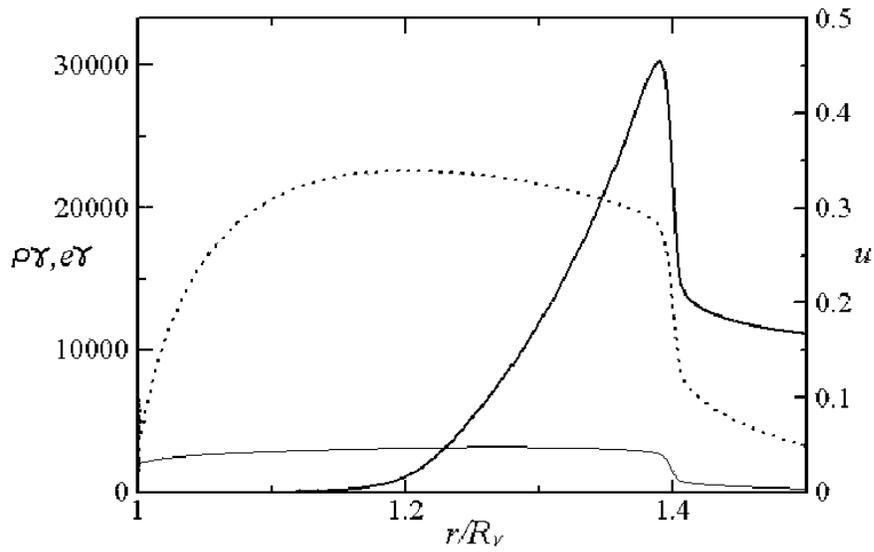}
\caption{Fireball at $\tau=1$ in the very dense case (see \S 3.2).
The initial matter density is $10^9$ g ${\rm cm^{-3}}$,
$T_{\rm eff}=15$ MeV, and $R_\nu=10$ km.
Here we include the effect of neutrino scattering.}
\end{figure}

\clearpage

\begin{figure}
\plotone{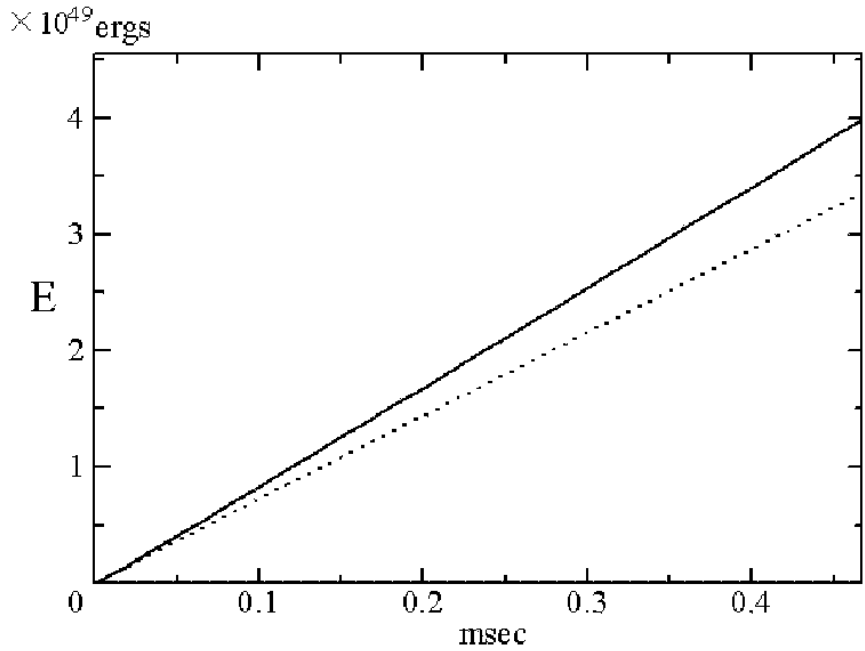}
\caption{Total energy deposited by neutrinos (solid line),
including the effect of neutrino scattering
and the energy deposited via neutrino pair-annihilation only (dotted line).
The physical parameters are the same as those in Fig. 9.}
\end{figure}

\clearpage

\begin{figure}
\plotone{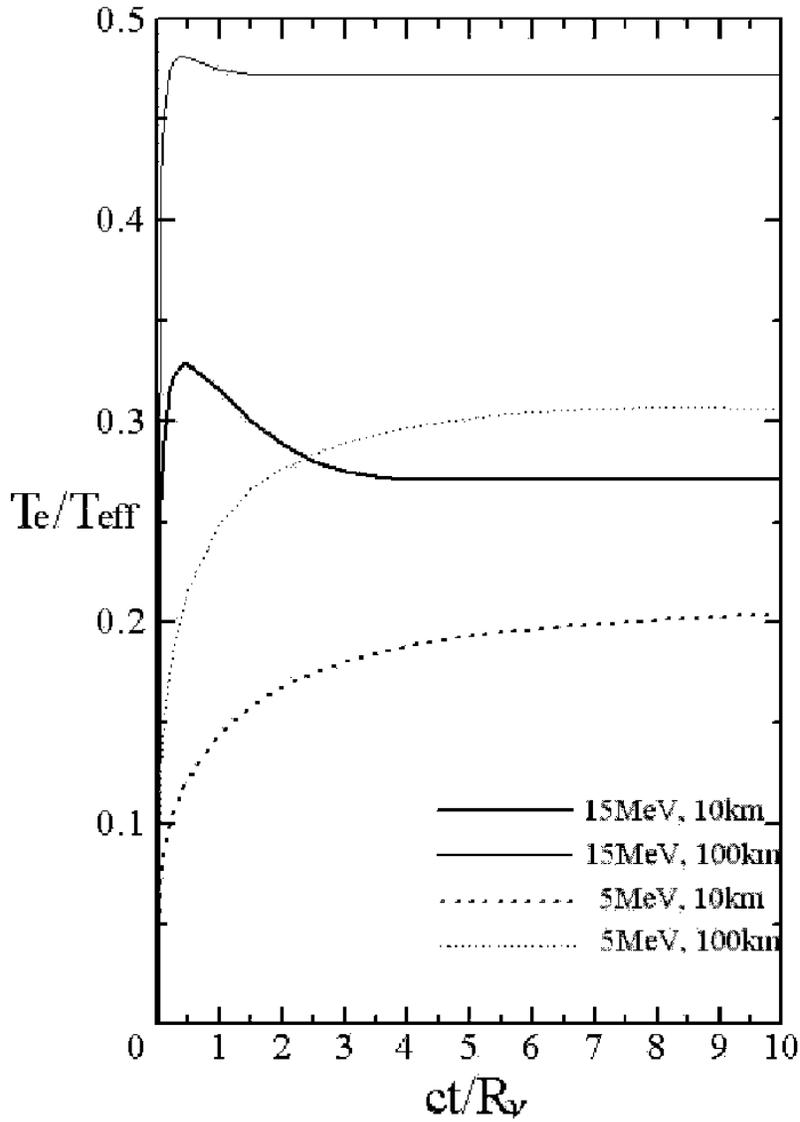}
\caption{Fireball temperature $T_{\rm e}$ at $r=R_\nu$.
The initial matter density is $10^9$ g ${\rm cm^{-3}}$
in all cases. The effective temperature $T_{\rm eff}$ and $R_\nu$ are
indicated.}
\end{figure}

\end{document}